\documentclass[aps,final,notitlepage,oneside,twocolumn,nobibnotes,nofootinbib,superscriptaddress,
noshowpacs,centertags,showkeys]{revtex4-1}

\usepackage[english]{babel}
\usepackage{graphicx}
\usepackage{latexsym}
\usepackage{amssymb}
\usepackage{amsmath}
\usepackage{psfrag}
\usepackage{float}
\usepackage[dvips]{color}

\begin{document}

\title{Nonlinear generation of sound waves in the pre-galactic epoch and 21 cm absorption}
\author{Yu. N. Eroshenko}\thanks{e-mail: eroshenko@inr.ac.ru}
\affiliation{Institute for Nuclear Research of the Russian Academy of Sciences
60th October Anniversary Prospect 7a, 117312 Moscow, Russia}

\date{\today}

\begin{abstract}
The structure of the sound waves generated in baryonic gas during the evolution of dark matter halos with masses less than the Jeans mass is calculated. In this case, the source of the gravitational field that creates the wave can be either at a linear stage (an evolving perturbation in dark matter) or at a nonlinear stage (detached and virialized objects). The peculiar velocity of baryons in the sound wave in the second order of velocity cause absorption of the relic radiation in the 21 cm line. It is shown that this additional absorption at sound waves ranges from fractions of a percent (at the redshifts $z\sim15-20$) to about percent (at $z\sim7-15$) of the absorption value in a homogeneous Universe, however, additional absorption may be larger in the case of a non-standard spectrum of small scale cosmological perturbations.
\end{abstract}


\maketitle 




\section{Introduction}

In recent years, the accuracy of observational data in cosmology has increased significantly and continues to improve. This applies both to the study of relic radiation and to the observation of early galaxies, quasars, absorption lines in gas and other processes. It can be expected that a number of new effects will need to be taken into account in order to quantify the data obtained.

Of great interest is the so-called Dark Ages cosmological epoch, at the end of which the first stars and galaxies appeared. The formation of gravitationally bound objects (galaxies, etc.) occurs during the growth of small dark matter (DM) perturbations with a transition eventually to nonlinear compression and virialization. The evolving medium consists of DM and baryon gas. Over time, both a general decrease in temperature and the growth of small inhomogeneities of inflationary origin occurred (we do not consider possible exotic seeds in the form of, for example, primordial black holes here). Before the redshift $z=z_1\simeq150$, the temperature of the baryon gas was kept at the level of the relic radiation temperature due to the Thomson scattering. At $z<z_1$ and before the epoch of reionization, the gas cooled adiabatically on average and moved in the gravitational fields of DM perturbations. In rare and dense DM objects with a mass greater than the Jeans mass, which entered the nonlinear stage early, baryons were captured and heated. These DM objects gave rise to the first stars and the first galaxies, whereas the most numerous perturbations of smaller scales could not capture baryons. It should be noted that recent observations of the J.~Webb  Space Telescope showed that there were more early galaxies than previously thought, which may be explained by the non-standard spectrum of perturbations at small scales \cite{Tkaetal23}.

The study of the baryon gas evolution in conjunction with the evolution of DM halos has been carried out in a variety of works, both by analytical and numerical methods. In the framework of linear theory, when all perturbations are considered weak, an analytical power-law solution was found in \cite{Pee84} for the epoch $z>z_1$, but only baryons within the DM condensation were considered, and their spreading was not taken into account. In most cases, objects with a mass greater than the Jeans mass were considered, and attention was paid only to gravitationally trapped baryons. This is due to the fact that on scales smaller than the Jeans length, perturbations in the baryonic gas run away in the form of sound waves and fade out, and therefore, as was usually thought, they are not of interest. In this paper, we show that despite the spreading of baryons, there are perturbations in the baryon gas density  at these scales, which at the level of percent can affect absorption in the 21~cm line. 

The growth of perturbations in dark matter and baryonic gas was previously described in linear theory only on the assumption that the magnitude of perturbations in both dark matter and baryonic gas is much less than one. In this paper, we investigate the process of generating sound waves, including the stage of virialization of the DM halo, i.e. when DM perturbations are highly nonlinear ($\delta_D>1$). In this case, the standard linearized equations for the evolution of perturbations in baryons are not valid in terms of the gravitational field source. Refined equations with a nonlinear source will be obtained and solved. Sound waves can cover quite large areas of space, significantly exceeding the radius of the DM object that generated it. Their amplitudes and prevalence depend on the statistics of DM gravitational clustering and the gas temperature in a particular epoch. 

Also, the absorption of relic radiation in sound waves in the 21~cm neutral hydrogen line is considered. The main part of the absorption effect is related to the collective velocity of the gas in the sound wave. Gas from a massive object spreads more slowly than in an undisturbed Hubble flow. It is known that the radial velocity gradient should lead to increased absorption in the lines. This effect was first indicated in \cite{Dubrovich77} and further substantiated in \cite{Zel78} in relation to absorption lines of molecules. In the work \cite{Dubrovich18}, it was hypothesized that the anomalous absorption observed by EDGES \cite{BowRogMon18} may be associated with the formation of the first structures in the Universe during the Dark Ages, and then in \cite{VasShc12,DubrovichGra19,DubrovichGraEro21} it was shown that regions of strong absorption can exist in proto-galaxies and on the periphery of the halo around supermassive black holes. However, such objects are very rare and cannot explain the full magnitude of the anomalous absorption detected by EDGES. Additional absorption due to the presence of peculiar velocities was also discussed in \cite{MesFurCen11,XuYueChe18}, and in \cite{XuYueChe18} the distribution of baryon gas outside the virial radius of the DM halo was also taken into account, however, consideration was limited only to massive objects (on scales larger than the Jeans mass, $\geq10^5M_\odot$), and the spreading of baryons (separation of baryon and DM fluxes), which in this paper we call a sound wave, was not considered. 

The calculations performed in this work indicate that the mechanism specified in \cite{Dubrovich18} takes place due to nonlinear effects in the second order in terms of gas velocities, whereas in a linear approximation, full compensation would occur. In addition, unlike \cite{Dubrovich18}, we consider the gas not in the DM objects themselves, but the extended sound waves created by these objects. The resulting correction to the absorption depth in a homogeneous gas turns out to be small, from fractions of a percent to several percent, but it can be expected that in the future the accuracy of observations will reach a level where such corrections can be seen. This will provide valuable information about low-mass DM halos (sources of sound waves) and, consequently, about the spectrum of cosmological perturbations on small scales. The shape of the spectrum, in turn, contains information about the inflationary potential responsible for the generating of perturbations.


\section{The structure of the sound wave}
\label{soundsec}

Let's consider a spherically symmetric perturbation of the gas density created by the DM condensation with a mass less than the Jeans mass. In this case, the thermal velocities of the gas atoms are greater than the escape velocity, the gas is not captured in the evolving DM object, but spreads out from it in the form of a sound wave. To calculate its structure, it is necessary to write down the Euler equations and the continuity equations for DM and baryons and linearize them. In \cite{Bon57}, an equation was obtained for the evolution of a spherical sound wave in a baryonic gas (without DM) in a linear approximation. We will do the same procedure for the two-component medium (DM and baryons). Unlike \cite{Bon57}, in the process of the equation deriving, we apply the linear approximation everywhere, except for the gravity source on the right-hand side of the equations. As a result, excluding velocities from the equations, we obtain 
\begin{equation}
\frac{\partial^2\delta_{\rm D}}{\partial t^2}+2H\frac{\partial\delta_{\rm D}}{\partial t}=4\pi G(\Delta\rho_{\rm D}+\Delta\rho_{\rm B}),
\label{eqdm}
\end{equation}
\begin{equation}
\frac{\partial^2\delta_{\rm B}}{\partial t^2}+2H\frac{\partial\delta_{\rm B}}{\partial t}=4\pi G(\Delta\rho_{\rm D}+\Delta\rho_{\rm B})+\frac{v_s^2}{a^2}\left(\frac{\partial^2\delta_{\rm B}}{\partial x^2}+\frac{2}{x}\frac{\partial\delta_{\rm B}}{\partial x}\right),
\label{eqbar}
\end{equation}
where $\delta_{\rm D}=\delta\rho_{\rm D}/\bar\rho_{\rm D}$ and $\delta_{\rm B}=\delta\rho_{\rm B}/\bar\rho_{\rm B}$ -- perturbations, respectively, in the DM density $\rho_{\rm D}$ and baryon gas $\rho_{\rm B}$. In the linear approximation, $\Delta\rho_{\rm D}=\delta_{\rm D}\bar\rho_{\rm D}$ with $\delta_{\rm D}\ll 1$, but in general, the sources of gravity $4\pi G(\Delta\rho_{\rm D}+\Delta\rho_{\rm B})$ can be non-linear and even can be created by other objects, for example, primordial black holes. The scale factor of the Universe is denoted by $a(t)$, and the Hubble constant is denoted by $H(t)$ (in the epoch under consideration, with good accuracy $a\propto t^{2/3}$, $H=2/(3t)$), $x=r/a(t)$ is the co-moving coordinate, $v_s=[5k_{\rm B}T(z)/(3m_p)]^{1/2}$ is the speed of sound in hydrogen (the contribution of helium is neglected), $T(z)$ is the gas temperature, $k_{\rm B}$ is the Boltzmann constant, and $m_p$ is the proton mass.

We assume that the sound wave in baryons is weak and is described by the equation (\ref{eqbar}), and instead of the equation (\ref{eqdm}) and the term $4\pi G\Delta\rho_{\rm D}$ in (\ref{eqbar}) we use the well-known nonlinear spherical symmetric solution for DM perturbation is the simplest spherical ``top-hat'' model. Thus, we consider the generation of a linear sound wave in baryons by a nonlinear DM object. For the best of our knowledge, the equation (\ref{eqbar}) has not been studied in this mode before.

In \cite{Pee84}, a linear solution was found for perturbations in a baryonic gas during the epoch $z>z_1\simeq150$, when Compton scattering on baryons of relic photons was strong, due to which the temperature in the gas was maintained at the level of the temperature of the relic radiation. At $z<z_1$, this condition is not fulfilled, and power-law solutions of the equations cannot be found. In addition, in \cite{Pee84} the solution is found in the form $\delta_{\rm B}=\xi\delta_{\rm D}$, i.e. for perturbations in baryons in the same place where there is a DM perturbation. In this paper, on the contrary, we will find an approximate solution at $z<z_1$ for spreading sound waves in the region beyond the contracting DM perturbation. 

We will consider the generation and evolution of sound waves in the epoch $7<z<z_1$, when the scattering of photons of microwave radiation on baryons became weak, the state of the baryon gas changes adiabatically, and its temperature $T\propto1/a(t)$. Since $\Delta\rho_{\rm B}\ll \Delta\rho_{\rm D}$, the influence of baryons on DM is neglected and in the sum of $\Delta\rho_{\rm D}+\Delta\rho_{\rm B}$ in the right-hand part of (\ref{eqbar}) we leave only $\Delta\rho_{\rm D}$.

We reduce the equation (\ref{eqbar}) to a wave equation with a constant velocity coefficient and with a source. To do this, we introduce the new time variable 
\begin{equation}
\eta=1-\frac{t_1^{1/3}}{t^{1/3}}
\end{equation}
and replace $s=\delta_{\rm B}/x$. Here and further, all values with the index ``1'' are taken at the redshift $z=z_1$, and the index ``eq'' marks the values taken at the time of transition from the radiation-dominated to the dust-like stage (equality moment) of the Universe evolution. As a result, we have the equation
\begin{equation}
\frac{\partial^2s}{\partial \eta^2}-w_s^2\frac{\partial^2s}{\partial x^2}=f(\eta,x),
\label{eqbareta}
\end{equation}
where  
\begin{equation}
w_s=\frac{3v_1t_1}{a_1^2},
\end{equation}
\begin{equation}
f(\eta,x)=\frac{6x}{(1-\eta)^2}\frac{\Delta\rho_{\rm D}}{\bar\rho_{\rm D}},
\end{equation}
and $v_1$ is the sound speed $v_s$ in the gas at the moment $t_1$. This velocity is derived from the dependence of the temperature of the baryon gas on the redshift, given, for example, in \cite{TseHir10}.
Let's denote 
\begin{equation}
r_s=3v_1t_1\simeq60\mbox{~pc,}
\end{equation}
and
\begin{equation}
M_s=\frac{4\pi}{3}r_s^3\rho_D(t_1)\simeq10^5M_\odot.
\end{equation}
The value $M_s$ plays the role of the Jeans mass in the epoch under consideration. Virialized DM objects with a mass of $M_D<M_s$ are not able to capture baryons due to the weak gravitational potential, and on the contrary, formed objects with a mass of $M_D>M_s$ (which we do not consider in this paper) retain gas, which leads to its heating.

We neglect the weak perturbations in the baryonic gas and their derivatives at the initial moment $t_1$. The solution of the wave equation (\ref{eqbareta}) is considered on a semi-direct $x\geq0$, and the boundary condition at zero $x=0$ has the form $\partial s/\partial x=0$. In this case, the solution of the boundary value problem is given by the d'Alembert formula, in which instead of the function $f(\eta,x)$, its even extension to the semidirect $x<0$ is used, i.e. $f_1(\eta,x)=f(\eta,x)$ for $x>0$ and $f_1(\eta,-x)=f_1(\eta,x)$. The solution is written as
\begin{equation}
s(\eta,x)=\frac{1}{2w_s}\int\limits_0^\eta d\tau\int\limits_{x-w_s(\eta-\tau)}^{x+w_s(\eta-\tau)}f_1(\xi,\tau)d\xi.
\label{dalamb}
\end{equation}
We will model the DM density profile by a Gaussian in the form
\begin{equation}
f_1(x,\eta)=x\gamma(\eta)e^{-\beta(\eta)x^2},
\end{equation}
then the solution (\ref{dalamb}) after integration by $\xi$ takes the form
\begin{eqnarray}
s(\eta,x)=\frac{3}{2w_s}\int\limits_0^\eta d\tau\frac{1}{(1-\tau)^2\beta(\tau)}\frac{\Delta\rho_{\rm D}(\tau)}{\bar\rho_{\rm D}(\tau)}
\nonumber
\\
\times\left[
e^{-\beta(\tau)[x-w_s(\eta-\tau)]^2}
-e^{-\beta(\tau)[x+w_s(\eta-\tau)]^2}
\right].
\label{daleqint}
\end{eqnarray}

The total DM mass in the object is
\begin{equation}
M_{\rm D}=\frac{4\pi}{3}\rho_{\rm D}R^3=\frac{4\pi}{3}\left(\rho_{\rm D}X^3\right)a^3,
\end{equation}
where $R=aX$ is the characteristic physical radius of the DM object, and the comoving radius is $X(\eta)=1/(\beta(\eta))^{1/2}$. 

Since $w_s=const$, the perturbation in the form of sound wave moves from the moment $\eta_i$ to the moment $\eta$ through the comoving distance $\Delta x=w_s(\eta-\eta_i)$. From here, one can find the physical displacement of the disturbance from the moment corresponding to the redshift $z_1$ to the moment corresponding to the current redshift $z$,
\begin{equation}
r_{\rm max}(z)=r_s\frac{1+z_1}{1+z}\left[\left(\frac{M_{\rm D}}{M_s}\right)^{1/3}+1-\frac{(1+z)^{1/2}}{(1+z_1)^{1/2}}\right].
\end{equation}
This value plays the role of the radius of propagation of the sound wave, if we ignore the weak perturbations that occurred at $z>z_1$. However, sound waves generated by the neighboring DM disturbances can overlap (this is taken into account later in the absorption calculation), forming a linear superposition of waves in the first order.

The exponent in (\ref{daleqint}) can be written as 
\begin{equation}
-\beta(\tau)w_s^2\left[\tau-\left(\eta-\frac{x}{w_s}\right)\right]^2,
\end{equation}
from where it can be seen that the main contribution to the density of the sound wave is made by the perturbations near a ``delayed'' moment of time
$\tau^*\simeq\eta-x/w_s$.

In the case of $\left(M_{\rm D}/M_s\right)^{1/3}\ll1$ at a great distance from the DM condensation (at $x\gg X$), an approximate solution can be written by performing the ``pass method'' integration in (\ref{daleqint}). At $\left(M_{\rm D}/M_s\right)^{1/3}\ll1$, the time of motion of the sound wave through the DM condensation is much less than the dynamic time of evolution of this condensation, therefore, the main change in values in (\ref{daleqint}) occurs due to variation of $\tau$ near $\tau^*$ in parentheses in the exponent, and you can get an approximate solution by taking the integral (\ref{daleqint}) provided that $\tau=\tau^*$ in all other places of the integrand.  As a result, we get
\begin{equation}
s(\eta,x)\simeq\frac{3\pi^{1/2}}{2w_s^2}\frac{1}{(1-\tau^*)^2\beta(\tau^*)}\frac{\Delta\rho_{\rm D}(\tau^*)X^3(\tau^*)}{\bar\rho_{\rm D}(\tau^*)},
\label{solsappr}
\end{equation}
where all values are taken at a delayed moment $\tau^*$. This DM object creates a sound wave that radiates from the center beyond the object itself, having the appearance of a baryon halo with increased density. The product $\Delta\rho_{\rm D}X^3$ is proportional to the value $D(\eta)=1-X^3/X_{\rm eq}^3$, the change of which over time can be approximately written as (if we neglect the duration of the transition to the nonlinear stage)
\begin{equation}
D(\eta)=\left\{
\begin{array}{lcr}
\delta_{\rm eq}(1+z_{\rm eq})/(1+z) & ; & z>z_c
\\
1  & ;  & z\leq z_c,
\end{array}
\right.
\label{dexpr}
\end{equation}
where the moment of transition to the nonlinear stage is $1+z_c=\delta_{\rm eq}(1+z_{\rm eq})$, and up to this point, evolution occurs according to a well-known linear law. The delayed moment of time (more precisely, the variable $\eta$) is given by the following expression
\begin{equation}
1-\tau^*=\frac{1+z}{1+z_1}\left[\frac{(1+z_1)^{1/2}}{(1+z)^{1/2}}+\frac{r}{r_s}\right]. 
\end{equation}
Then the solution for the expanding sound wave $\delta_B=s/x=sa(\eta)/r$ can be written in an approximate form
\begin{equation}
\delta_B=\left\{
\begin{array}{l}
\frac{3\pi^{1/2}}{2}\left(\frac{r}{r_s}\right)^{-1}\left[\frac{(1+z_1)^{1/2}}{(1+z)^{1/2}}+\frac{r}{r_s}\right]^{-4}\frac{M_{\rm D}}{M_s} 
\\
\times\frac{\delta_{\rm eq}(1+z_{\rm eq})(1+z_1)^4}{(1+z)^5}; \mbox{~at~} z>z_c \mbox{~or ~} z<z_c, r>r_c
\\
\\
\frac{3\pi^{1/2}}{2}\left(\frac{r}{r_s}\right)^{-1}\left[\frac{(1+z_1)^{1/2}}{(1+z)^{1/2}}+\frac{r}{r_s}\right]^{-2}\frac{M_{\rm D}}{M_s} 
\\
\times\frac{(1+z_1)^3}{(1+z)^3}; \mbox{~at~} z<z_c, r<r_c,
\end{array}
\right.
\label{apprsol}
\end{equation}
where $r_c=r_{\rm max}(z_c)$. Thus, we obtained the density distribution in the sound wave for both linear and nonlinear modes of the evolution of the DM object, and the transition along the radius between the two expressions in (\ref{apprsol}) is associated with the transition of the central object to a nonlinear stage of evolution. 

The calculation of the attenuation decrement in gas shows that in the Dark Ages epoch, the conditions are such that the sound wave, before weakening by $e$ times, manages to spread over a distance at least slightly exceeding the radius of the DM object that gave rise to it. Indeed, the collision cross section of hydrogen atoms $\sigma_{\rm coll}=4\pi a_B^2$, where the Bohr radius $a_B=\hbar^2/(m_ee^2)=0.53\times10^{-8}$~cm. From here we get the free path length $\lambda=1/(n\sigma_{\rm coll})\simeq1.8\left((1+z)/16\right)^{-3}\mbox{~pc}$, where $n$ is the number density of the hydrogen atoms. The length of the sound wave has the order of size $R$ of the DM condensation. At the stage of linear evolution $R(z)\simeq r_s\left(M_{\rm D}/M_s\right)^{1/3}(1+z_1)/(1+z)$. As is known from the theory of sound waves in a gas, the wave propagates over a distance of $\sim R(R/\lambda)$ before its attenuation by $e$ times. Numerically, we have 
\begin{equation}
\frac{R}{\lambda}\simeq300\left(\frac{1+z}{16}\right)^2\left(\frac{M_{\rm D}}{M_s}\right)^{1/3}.
\end{equation}
This means that in the cases we are considering, when $\left(M_{\rm D}/M_s\right)^{1/3}\sim1$, attenuation is not important, and sound waves can cover quite large areas of space. Their amplitudes and prevalence depend on the statistics of DM gravitational clustering  and the temperature of the gas in a particular epoch.

The expressions (\ref{apprsol}), although they provide a qualitative illustration of the structure of the sound wave, for $\left(M_{\rm D}/M_s\right)^{1/3}\sim1$ have low accuracy, since in this case the propagation time of the sound wave within the object size $r\leq r_s\left(M_{\rm D}/M_s\right)^{1/3}(1+z_1)/(1+z)$ plays a large role. From the point of view of absorption in the 21 cm line, we are most interested in objects with $\left(M_{\rm D}/M_s\right)^{1/3}\sim1$, which in their central region create a significant gradient of the radial velocity of baryons. Therefore, we do not consider approximate expressions (\ref{apprsol}) further, but apply the exact solution (\ref{daleqint}), performing numerical integration in it. When integrating, it is convenient to switch from the variable $\tau$ to the redshift $z$. In addition, we take into account the evolution of DM condensation at the linear and subsequent nonlinear stages within the framework of the spherical ``top-hat'' model. In this model, the evolution of the condensation radius over time is expressed in the parametric form
\begin{equation}
\left\{
\begin{array}{l}
R=R_{\rm eq}\frac{3}{5\delta_{\rm eq}}\cos^2\theta,
\\
\\
t_s/t_{\rm eq}=\frac{3\pi}{4}\left(\frac{5\delta_{\rm eq}}{3}\right)^{-3/2},
\\
\\
\theta+\frac{1}{2}\sin(2\theta)=\frac{2}{3}\left(\frac{5\delta_{\rm eq}}{3}\right)^{3/2}\frac{t-t_s}{t_{\rm eq}},
\\
\\
\theta_{\rm eq}=-\frac{\pi}{2}+\left(\frac{5\delta_{\rm eq}}{3}\right)^{1/2},
\end{array}
\right.
\label{sphsol}
\end{equation}
where
\begin{equation}
R_{\rm eq}=\left(\frac{3M_D}{4\pi\rho_{\rm eq}}\right)^{1/3},
\label{reqeq}
\end{equation}
and the parameter $\theta$ changes from $\theta_{\rm eq}$ (at the moment $t_{\rm eq}$) to $\pi/4$ when the radius is halved from the maximum value. Near this moment, the DM condensation is virialized, and the radius does not change further, but the virialized object continues to exert a gravitational influence on the baryonic gas, slowing its spreading. There is a connection
\begin{equation}
\frac{R_{\rm eq}}{r_s}=\left(\frac{M_{\rm D}}{M_s}\right)^{1/3}\frac{1+z_1}{1+z_{\rm eq}}.
\label{reqeq}
\end{equation}
The redshift $z$ is associated with the time $1+z\propto t^{-2/3}$, and the corresponding value of the parameter $\theta$ and the ratio $R/R_{\rm eq}$ is found by numerically calculating the root of the equation in (\ref{sphsol}). Also from (\ref{sphsol}) we find 
\begin{equation}
\frac{\Delta\rho_D}{\bar\rho_D}=\frac{X_{\rm eq}^3}{X^3}-1=\left\{
\begin{array}{l}
\left(\frac{5\delta_{\rm eq}}{3\cos^2\theta}\right)^3\frac{(1+z_{\rm eq})^3}{(1+z)^3}-1; z>z_v
\\
\\
\left(\frac{10\delta_{\rm eq}}{3}\right)^3\frac{(1+z_{\rm eq})^3}{(1+z)^3}-1; z\leq z_v,
\end{array}
\right.
\label{xxeq}
\end{equation}
entering in (\ref{daleqint}). Here $z_v$ corresponds to the moment of virialization at $\theta=\pi/4$ and is also found numerically. For $z\gg z_v$, the expression (\ref{xxeq}) reproduces the well-known solution for the linear stage of evolution $\delta_D\propto t^{2/3}$. 
 
\begin{figure}
	\begin{center}
\includegraphics[angle=0,width=0.45\textwidth]{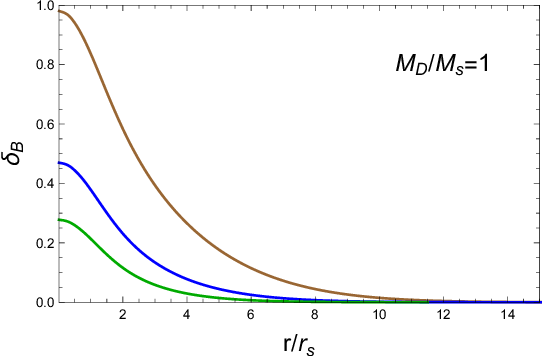}
	\end{center}
\caption{Density perturbations in sound waves for $M_D/M_s=1$ and $z=$10, 15, 20 (from top to bottom).}
	\label{grdel1}
\end{figure}

\begin{figure}
	\begin{center}
\includegraphics[angle=0,width=0.45\textwidth]{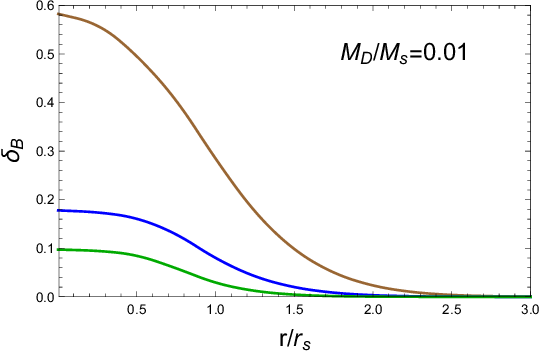}
	\end{center}
\caption{Density perturbations in sound waves for $M_D/M_s=0.01$ and $z=$10, 15, 20 (from top to bottom).}
	\label{grdel001}
\end{figure}

The resulting perturbation in the sound wave is shown in Fig.~\ref{grdel1} and Fig.~\ref{grdel001} for DM density perturbations with a mean square value $\delta_{\rm eq}\simeq\sigma_{\rm eq}(M_D)$. The primary spectrum of these perturbations is taken to be power-law with normalization to the Planck data.


\begin{figure}
	\begin{center}
\includegraphics[angle=0,width=0.45\textwidth]{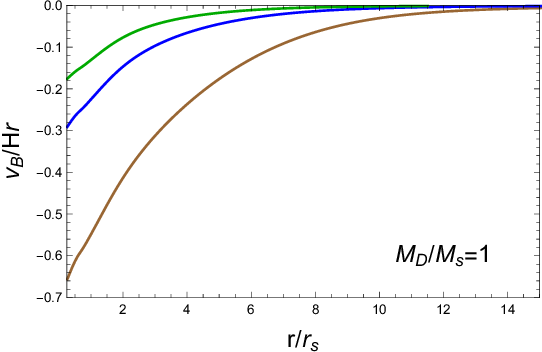}
	\end{center}
\caption{The peculiar velocity in sound waves with respect to the Hubble velocity for $M_D/M_s=1$ and $z=$20, 15, 10 (from top to bottom).}
	\label{grv1}
\end{figure}

\begin{figure}
	\begin{center}
\includegraphics[angle=0,width=0.45\textwidth]{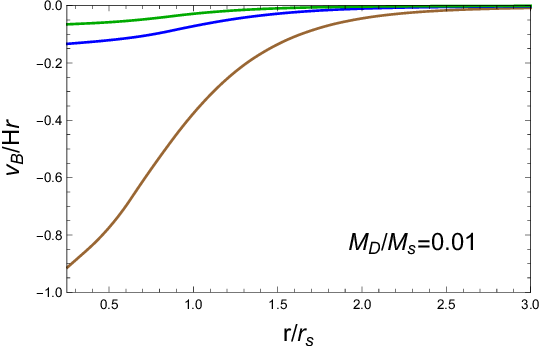}
	\end{center}
\caption{The peculiar velocity in sound waves with respect to the Hubble velocity for $M_D/M_s=0.01$ and $z=$20, 15, 10 (from top to bottom).}
	\label{grv001}
\end{figure}

Now let's find the peculiar velocities in the sound wave spreading from the center. Knowing $\delta_{\rm B}(z,r)$, the velocity can be found from the continuity equation in the form
\begin{equation}
v_{\rm B}=-\frac{1}{\bar\rho_{\rm B}r^2}\int\limits_0^r drr^2\left[\frac{\partial\delta\rho_{\rm B}}{\partial t}+3H\delta\rho_{\rm B}+Hr\frac{\partial\delta\rho_{\rm B}}{\partial r}\right]+\frac{\tilde F(t)}{r^2},
\end{equation}
where $\tilde F(t)$ is a function defined by boundary conditions, and for finiteness of velocity at $r\to 0$ it is necessary to put $\tilde F(t)=0$. After integrating by parts the term with $\partial\delta\rho_{\rm B}/\partial r$ and transformations, one can write
\begin{equation}
\frac{v_{\rm B}}{Hr}=-\delta_{\rm B}+\frac{(1+z)}{y^3}\frac{\partial}{\partial z}\int\limits_0^y dy'y'^2\delta_{\rm B}+\frac{3}{y^3}\int\limits_0^y dy'y'^2\delta_{\rm B},
\label{velhr}
\end{equation}
where $y=r/r_s$. The value (\ref{velhr}) is shown in Fig.~\ref{grv1} and Fig.~\ref{grv001}.


\section{Absorption in the 21 cm line}
\label{physcondsec}

When calculating the absorption in the 21 cm line on sound waves, first we find the absorption in a single sound wave propagating from one object, and then we perform statistical averaging by taking into account the proportion of the volume of space occupied by the waves. We assume that the spin temperature $T_s$ in neutral hydrogen is maintained at the level of the kinetic temperature of the gas due to the Wouthuysen–Field effect (see the review \cite{FurOhBri06}). The absorption value in the line is 21 cm \cite{FurOhBri06}
\begin{eqnarray}
\delta T_b&\simeq&27x_{HI}(1+\delta_B)\left(\frac{H}{\partial v_B/\partial s+H}\right)\left(1-\frac{T_\gamma}{T_s}\right)
\nonumber
\\
&\times&\left(\frac{1+z}{10}\frac{0.15}{\Omega_mh^2}\right)\left(\frac{\Omega_Bh^2}{0.0023}\right)\mbox{~mK},
\label{dtfin}
\end{eqnarray}
where $x_{HI}$ is the fraction of neutral hydrogen, $T_\gamma$ is the temperature of the relic radiation, $\Omega_m$ and $\Omega_B$ are cosmological parameters of the DM density and baryons, respectively, $h$ is the value of the Hubble constant in units of 100~km~s$^{-1}$Mpc$^{-1}$.

To calculate, we need to estimate the magnitude of the peculiar velocity gradient along the line-of-sight
\begin{equation}
\frac{\partial v_B}{\partial s}=\frac{\partial v_B}{\partial r}\cos\phi. 
\end{equation}
It is important to note that in the case of spherical symmetry, spatial regions with the same module $\cos\phi$, but with different signs, are equally common, therefore, when integrating over a spherical sound wave, $\partial v_B/\partial s$ in these two types of regions enters with the same weight factor $(1+\delta_{\rm B})$. Due to this, the contributions of density perturbations and velocity gradients in the second order of velocity are not entangled with each other and enter as two separate terms. This conclusion holds for the ellipsoidal distribution as well. In the presence of non-spherical quadrupole type shape, the situation becomes more complicated, and additional corrections may arise, but we do not consider this case in this paper. 

For evaluation, we will consider only cosmological perturbations of the DM density with the mean square value $\delta_{\rm eq}\simeq\sigma_{\rm eq}(M_D)$. In this case, the Press--Schechter theory cannot be directly applied, because we consider not only virialized objects, but also those that are still at a linear stage of evolution (a more accurate calculation, taking into account the distribution of all objects, is much more complicated and is planned in future works). The calculations described below have shown that objects with $M_D\sim0.05 M_s$ make the greatest contribution to absorption in the epoch under consideration before the reionization of the Universe, since they are quite numerous, just enter the nonlinear stage and create noticeable gradients of peculiar velocities.  In the case of a Gaussian distribution, half (in terms of volume) of perturbations is positive $\delta_{\rm eq}>0$, and half is negative $\delta_{\rm eq}<0$. The sound waves we are considering are created by positive perturbations, so half of the average distance between neighboring DM clusters with masses $M_D$ is 
\begin{equation}
l(z)=\frac{1}{2}\left(\frac{0.5\rho_{\rm eq}(1+z_{\rm eq})^3}{M_D(1+z)^3}\right)^{-1/3}.
\end{equation}
Let's denote 
\begin{equation}
L(z)=min\{l(z),r_{\rm max}(z)\},
\end{equation}
then the average over volume can be written as
\begin{eqnarray}
&~&\left<\frac{(1+\delta_{\rm B})H}{\partial v_B/\partial s+H}\right>=
\label{srednpogl}
\\
&=&1+\frac{1}{l^3(z)}\int\limits_0^{L(z)}drr^2\left(3\delta_{\rm B}(r)+\frac{1}{2H^2}\left(\frac{\partial v_B}{\partial r}\right)^2\right),
\nonumber
\end{eqnarray}
where, in the integrand, the $\phi$ angle integration has already been performed. The first term in (\ref{srednpogl}) corresponds to absorption in a homogeneous universe. The value of $\delta_{\rm B}(r)$ in parentheses under the integral corresponds to absorption fluctuations due to the presence of density inhomogeneities. Due to the conservation of the number of hydrogen atoms, in total in the regions with $\delta_{\rm B}>0$ and $\delta_{\rm B}\leq 0$, density inhomogeneities are exactly compensated for averaging over the entire volume of the Universe, therefore only the last term under the integral associated with the gradient of the peculiar speeds, and only this we will take into account further.

\begin{figure}
	\begin{center}
\includegraphics[angle=0,width=0.45\textwidth]{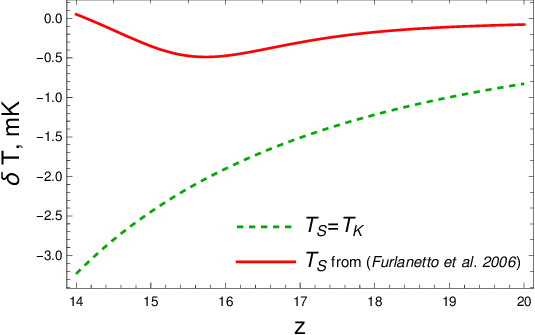}
	\end{center}
\caption{The absorption depth produced by objects with $M_D\sim0.05 M_s$ in the 21~cm line, depending on the redshift $z$. The upper continuous
curve shows the absorption for the spin temperature calculated in \cite{FurOhBri06}. The lower dashed curve corresponds to the case when the spin temperature is equal to the kinetic temperature of the gas.}
\label{brightfig}
\end{figure}

The result of the numerical calculation of (\ref{dtfin}) is shown at Fig.~\ref{brightfig}.
For comparison, we point out that in a homogeneous Universe, the absorption depth is $(\delta T)^{\rm hom}\simeq(-200\div-230)$~mK, and in the EDGES observations an anomalous absorption of $\delta T\simeq -500^{+200}_{-500}$~mK (at 99\% confidence level) at the redshift $z\simeq17$ was obtained  \cite{BowRogMon18}. Thus, the absorption on sound waves created by DM objects with masses $M_D\leq M_s$ ranges from fractions of a percent (at $z\sim15-20$) to several percent (at $z\sim7-15$) of the total absorption value in the case of a homogeneous Universe.

Objects with $M>M_s$ are quite rare, but on the periphery of such objects there should be areas of strong absorption \cite{VasShc12,DubrovichGraEro21}. Absorption occurs most effectively in the area of stopping the expansion of DM layers. The perturbation theory method used in this work does not allow us to consider such objects. Absorption for objects with $M>M_s$ should slightly increase the depth of total absorption.


\section{Conclusion}

In this paper, a new solution has been found for perturbations of the density and velocity field of sound waves in a baryonic gas created by evolving DM objects. The transition from the linear stage of the DM objects evolution to the nonlinear one is traced. With the help of the obtained solution, the absorption of relic photons in the 21 cm line was investigated. The absorption effect generally corresponds to the mechanism considered in \cite{Dubrovich77,Zel78,Dubrovich18,DubrovichGra19} and associated with the radial velocity gradient, but in our case it acts in the second order of velocity due to compensation of the first order terms. It is found that the absorption depth ranges from fractions of a percent to several percent of the absorption depth in a homogeneous Universe. Thus, this effect is not vanishingly small. It is possible that the accuracy of cosmological observations will reach this level in the future, and in order to fully describe the processes in the early Universe, sound waves on scales smaller than the Jeans length will have to be taken into account. 

Although objects with $M<M_s$ do not capture baryons and do not even completely stop the Hubble flow, they create sound waves around themselves, on which increased absorption occurs.  The strongest absorption in sound waves occurs when the characteristic mass in the Press--Schechter mass function is compared with the mass of $\sim(0.01-1)M_s$. In this case, there is an optimal ratio between the absorption depth and the number of DM objects that create sound waves.

The absorption on the sound waves may be greater if the spectrum of cosmological perturbations had an additional excess at small scales. In the work \cite{Tkaetal23} it is shown that due to such an excess, the appearance of early galaxies observed by the J.~Webb Space Telescope can be explained. 

At present time ($z\leq 1$), the solutions we have obtained for sound waves are not applicable, but it can be expected that the considered regions of perturbation of baryons have evolved by now into weak inhomogeneities of intergalactic gas with sizes of several kpc, overlapping with each other.



\begin{thebibliography}{99}

\bibitem{Tkaetal23} M. V. Tkachev, S .V. Pilipenko, E. V. Mikheeva, V. N. Lukash, arXiv:2307.13774 [astro-ph.CO].

\bibitem{Pee84} P. J. E. Peebles, Astrophys. J. 277, 470 (1984).

\bibitem{Dubrovich77} V. K. Dubrovich, Soviet Astronomy Letters 3, 128 (1977).

\bibitem{Zel78} Y. B. Zel’dovich, Soviet Astronomy Letters 4, 88 (1978).

\bibitem{Dubrovich18} V. K. Dubrovich, arXiv:1805.04430 [astro-ph.CO].

\bibitem{BowRogMon18} J. D. Bowman, A. E. E. Rogers, R. A. Monsalve, T. J. Mozdzen, and N. Mahesh, Nature 555, 67 (2018).

\bibitem{VasShc12} E. O. Vasiliev, Yu. A. Shchekinov, Astronomy Reports 56, 77 (2012).

\bibitem{DubrovichGra19} V. K. Dubrovich, S. I. Grachev, Astron. Lett. 45, 701 (2019).

\bibitem{DubrovichGraEro21} V. K. Dubrovich, Yu. N. Eroshenko, S. I. Grachev, Monthly Notices of the Royal Astronomical Society 503, 3081 (2021).

\bibitem{MesFurCen11}  A.  Mesinger, S. Furlanetto, R. Cen, Monthly Notices of the Royal Astronomical Society 411, 955
(2011).   
    
\bibitem{XuYueChe18} Y. Xu, B. Yue, X. Chen, Astrophys. J. 869, 42 (2018).  

\bibitem{Bon57} W. B. Bonnor, Monthly Notices of the Royal Astronomical Society 117, 104 (1957).

\bibitem{TseHir10} D. Tseliakhovich, C. Hirata, Phys. Rev. D 82, 083520 (2010).

\bibitem{FurOhBri06} S. R. Furlanetto,  S. P. Oh, F. H. Briggs, Physics Reports 433, 181 (2006).

\end{thebibliography}
\end{document}